 \documentclass[final,3p,times,twocolumn]{elsarticle}
\usepackage{amssymb}
\journal{Nuclear Physics B}

\begin{document}

\begin{frontmatter}



\title{New dynamical pair breaking effect}


\author{M. Mirea }

\address{Horia Hulubei National Institute for Physics and Nuclear Engineering,
P.O. Box MG-6, 077125 Bucharest-Magurele, Romania}

\begin{abstract}

A dynamical pair breaking effect is evidenced at very low excitation
energies. For this purpose, a new set of
time-dependent coupled channel equations for pair-breaking in superfluid
systems are deduced from the variational principle.
These equations give the probability to destroy
or to create a Cooper pair under
the action of some perturbations or
when the mean field varies in time.
 The odd-even effect in fission
is investigated within the model as an example.
For this purpose, the time-dependent probability
to find the system in a seniority-one or in a seniority-two state is
restricted in the sense that the perturbations are considered
only in the avoided crossing regions.

\end{abstract}


\end{frontmatter}


\section{Introduction}
\label{ntro}

In atomic or nuclear molecules, fermions move in a common field generated by several
centers of force. The potential depends on the so-called generalized coordinates.
The most used generalized variables are function of the inter-nuclear distances and
their angular dependencies. In such a molecular description, the motion of the 
nuclear centers is considered adiabatically slow compared with the
rearrangement of the mean field. The 
atoms or the nuclei share their valence electrons and their outer-bound
nucleons, respectively. Enhancements and structures in the energy dependence
of the cross section in atomic \cite{fano,licten} and nuclear \cite{milek,park}
reactions reflect a
special signature of the presence of molecular orbitals. The behavior of the
cross section is usually explained in terms of the Landau-Zener effect.

The Landau-Zener effect \cite{landau,zener} describes the non-adiabatic
transitions at avoided crossing regions between potential 
curves \cite{marcus,wheeler}
or energy levels \cite{rubb,mirea}. Two levels with the same good
quantum numbers associated to some symmetries cannot intersect 
and exhibit avoided level crossing regions. A small perturbation energy
is always available in these regions, being responsible for
the level slippage mechanism. Recently, 
the time-dependent equations
that describe quantitatively the Landau-Zener promotion mechanism were
generalized for molecular potentials that include pairing
residual interactions \cite{mirea2}. As evidenced below,
this generalization also provides a way to formulate a theory
for a new dynamical pair breaking mechanism.
Originally, the pairing formalism, reflected in the BCS equations, 
was used to explain superconducting states
in terms of two electrons pairs with opposite momenta and spin
near the Fermi surface \cite{bcs}.

\section{Formalism}
\label{formal}

 The dynamical pair breaking effect emerges
from a new set of coupled channel equations  deduced for 
 the time-dependent
probability to find the system in a seniority-one state 
or in a seniority-two one.
The variation
of the mean field
is considered slow enough, so that fermions follow eigenstates of the 
instantaneous mean field. 
In such an approximation, if the interactions produced
in the avoided crossing regions or those due to the 
Coriolis coupling are not taken into consideration, other 
perturbations between two different states are not possible.
In the following, the calculations are restricted only
for perturbations produced in the avoided crossing regions.
This approximation does not affect the essential features 
of the model but leads to a considerable simplification
of the mathematical apparatus.

 The starting point is
a many-body 
Hamiltonian with pairing residual interactions. This Hamiltonian
 depends on some time-dependent collective parameters
$q(t)=\{q_{i}(t)\}$ ($i=1,...n$), such as the internuclear 
distances between atoms or nuclei: 
\begin{equation}
H(t)=\sum_{k>0} \epsilon_{k}[q(t)](a_{k}^{+}a_{k}+a_{\bar k}^{+}a_{\bar k})-G\sum_{k,i>0}a_{k}^{+}a_{\bar{k}}^{+}
a_{i}a_{\bar{i}}.
\label{ham1}
\end{equation}
Here, $\epsilon_{k}$ are single-particle energies of the molecular potential,
 $a_{k}^{+}$ and $a_{k}$ denote operators for creating
and destroying a particle in the state $k$, respectively. The state characterized
by a bar signifies the time-reversed partner of a pair.
The pairing correlation arise from the short range interaction 
between fermions moving in time-reversed orbits. The
essential feature of the pairing correction can be described in terms
of a constant pairing interaction $G$ acting between a given number of
 particles. In this paper, the sum over pairs generally runs over
the index $k$. Because the pairing equations diverge for
an infinite number of levels, a limited number of levels are used
in the calculation, that is $N$ levels above and below the
Fermi energy $E_{F}$.

\begin{figure}
\begin{center}
\resizebox{0.50\textwidth}{!}{
  \includegraphics{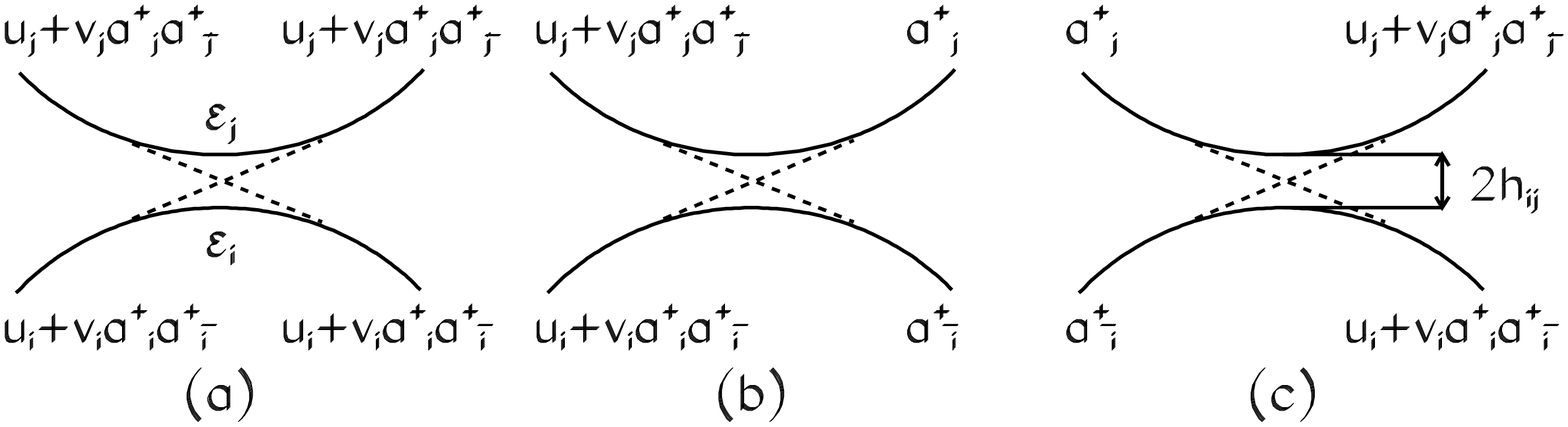}}
\caption
{
 Ideal avoided crossing regions between two adiabatic levels 
$\epsilon_i$ and $\epsilon_j$. Three
possible transitions between configurations
in an avoided crossing region in the superfluid model are
displayed: (a) The configuration
remains unchanged after the passage through the avoided crossing region.
(b) A pair is broken. (c) A pair is created.
}
\label{figura1}
\end{center}
\end{figure}

Using quasiparticle creation and annihilation operators 
\begin{eqnarray}
\alpha_{k(\gamma)}&=&u_{k(\gamma)}a_{k}-v_{k(\gamma)}a_{\bar k}^{+};\nonumber\\
\alpha_{\bar{k}(\gamma)}&=&u_{k(\gamma)}a_{\bar k}+v_{k(\gamma)}a_{ k}^{+};\\
\alpha_{k(\gamma)}^{+}&=&u_{k(\gamma)}a_{k}^{+}-v_{k(\gamma)}^{*}a_{\bar k};\nonumber\\
\alpha_{\bar{k}(\gamma)}^{+}&=&u_{k(\gamma)}a_{\bar k}^{+}+v_{k(\gamma)}^{*}a_{ k};\nonumber
\label{anho2}
\end{eqnarray}
it is possible to construct some interactions able  to break a Cooper
pair when the system traverses a avoided crossing region.
The parameters $v_{k(\gamma)}$ and $u_{k(\gamma)}$ are occupation and vacancy amplitudes,
respectively, for a pair occupying the single-particle level $k$ of 
the configuration $(\gamma)$.
The seniority-zero configuration is labeled with $(\gamma)=(0)$ and the
seniority-two configuration by a pair of indexes denoting the
levels blocked by the unpaired fermions $(\gamma)=(ij)$.
The three situations plotted in Fig. \ref{figura1}
can be modeled within products of
such creation and annihilation operators acting on
Bogoliubov wave functions. In the plot \ref{figura1}(a), the Cooper pairs remain
on the adiabatic levels $\epsilon_{i}$ and $\epsilon_{j}$ after the passage
through the avoided crossing region,  
in Fig. \ref{figura1} (b) the pair destruction is
illustrated,
 while in Fig. \ref{figura1} (c) two fermions generate a pair after the passage
through an avoided crossing region.  Formally, to describe
these three situations, an interaction in the avoided crossing can be postulated
as follows:
\begin{eqnarray}
H'(t)&=&\sum_{i,j\ne i}^{n}h_{ij}[q(t)]
\left[\alpha_{i(0)}\alpha_{\bar{j}(0)}
\prod_{k\ne i,j}\alpha_{k(0)}a_{k}^{+}a_{k}\alpha_{k(ij)}^{+}\right.\nonumber\\
& & \left. +\alpha_{i(0)}^{+}\alpha_{\bar{j}(0)}^{+}
\prod_{k\ne i,j}\alpha_{k(ij)}a_{k}^{+}a_{k}\alpha_{k(0)}^{+}\right]
,
\label{cort}
\end{eqnarray}
where $h_{ij}$ is the interaction between the levels. 
The form of the perturbation (\ref{cort}) was postulated in Ref.
\cite{mirea2} and was successfully used to generalize the Landau-Zener
effect in seniority-one superfluid systems. Acting on a suited
Bogoliubov wave function, the product over $k$
transforms the seniority-two configuration in the seniority-zero one 
in the case of
the first term in the left hand of Eq. (\ref{cort}), and
vice-versa in the case of the second term. 
If the product $\alpha_{i(0)}\alpha_{\bar{j}(0)}$ 
acts on a seniority-zero function, then it annihilates a pair
and creates two unpaired fermions in states $i$ and $\bar{j}$. If the
product $\alpha_{i(0)}^{+}\alpha_{\bar{j}(0)}^{+}$ acts on a seniority-two
wave function, then it creates a pair distributed on both orbitals $i$ and $j$.
In order to obtain the equations of motion, we shall start from the variational
principle taking the following energy functional
\begin{eqnarray}
{\cal{L}}=<\varphi \mid H-i\hbar{\partial\over \partial t}+H'-\lambda \hat{N}\mid \varphi >
\label{expr0}
\end{eqnarray}
and by assuming the many-body state formally expanded as a superposition of
time dependent BCS seniority-zero and seniority-two adiabatic wave functions
\begin{eqnarray}
\mid\varphi(t)>&=&\left[c_{0}(t)\prod_{k}
\left(u_{k(0)}(t)+v_{k(0)}(t)a_{k}^{+}a_{\bar{k}}^{+}\right)\right.\nonumber\\
 & & +\sum_{j,l\ne j}c_{jl}(t)a_{j}^{+}a_{\bar{l}}^{+}\\
& & \left.\times\prod_{k\ne j,l}
\left(u_{k(jl)}(t)+v_{k(jl)}(t)a_{k}^{+}a_{\bar{k}}^{+}\right)\right]\mid 0>,\nonumber
\label{wf2}
\end{eqnarray}
where $c_{0}$ and $c_{jl}$ are amplitudes of the
two kind of configurations. Here, $\lambda$ is the chemical potential, and 
$\hat{N}$ is the particle number operator. To minimize this functional, the
expression (\ref{expr0}) is derived with respect the independent
variables $v_{k(0)}$, $v_{k(jl)}$, 
$c_{0}$,  $c_{jl}$,  together with their complex conjugates,
 and the resulting equations
are set to zero.
Eventually, eight coupled-channel equations are obtained:
\begin{eqnarray}
i\hbar \dot{\rho}_{k(0)}&=&\kappa_{k(0)}\Delta_{0}^{*}-
\kappa_{k(0)}^{*}\Delta_{0};\nonumber\\
i\hbar \dot{\rho}_{k(jl)}&=&\kappa_{k(jl)}\Delta_{jl}^{*}-
\kappa_{k(jl)}^{*}\Delta_{jl};\nonumber\\
i\hbar \dot{\kappa}_{k(0)}&=&\left(2\rho_{k(0)}-1\right)\Delta_{0}+
2\kappa_{k(0)}\left(\epsilon_{k}-\lambda_{0}\right)\nonumber\\
 & & -2G\rho_{k(0)}\kappa_{k(0)};\nonumber\\
i\hbar \dot{\kappa}_{k(jl)}&=&\left(2\rho_{k(jl)}-1\right)\Delta_{jl}+
2\kappa_{k(jl)}\left(\epsilon_{k}-\lambda_{jl}\right)\nonumber\\
& & -2G\rho_{k(jl)}\kappa_{k(jl)};\nonumber\\
i\hbar \dot{P}_{0}&=&\sum_{l,j\ne l}h_{jl}(S_{0jl}^{*}-S_{0jl});\label{system}\\
i\hbar \dot{P}_{jl}&=&h_{jl}(S_{0jl}-S_{0jl}^{*});\nonumber\\
i\hbar \dot{S}_{0jl}&=&S_{0jl}(E_{0}-N\lambda_{0}-E_{jl}+N\lambda_{jl})\nonumber\\
& & +S_{0jl}\left(\sum_{k\ne j,l}T_{k(jl)}-\sum_{k}T_{k(0)}\right)\nonumber\\
& & +\sum_{m,n\ne m}h_{mn}S_{mnjl}+h_{jl}(P_{jl}-P_{0});\nonumber\\
i\hbar \dot{S}_{mnjl}&=&S_{mnjl}(E_{mn}-N\lambda_{mn}-E_{jl}+N\lambda_{jl})\nonumber\\
& & +
S_{mnjl}\left(\sum_{k\ne m,n}T_{k(mn)}-\sum_{k\ne j,l}T_{k(jl)}\right)\nonumber\\
& & +h_{mn}S_{0jl}-h_{jl}S_{0mn}^{*};\nonumber
\end{eqnarray}
where the partial derivatives with respect the time are denoted by a dot.
The sums are restricted by the conditions
$j\ne l$, $m\ne n$, $m\ne j$, and $n\ne l$. $E_{\gamma}$ are exactly
the expected values of the Hamiltonian (\ref{ham1}) for
the seniority-zero or seniority-two 
configurations:
\begin{eqnarray}
E_{0}&=&2\sum_{k} \rho_{k(0)}\epsilon_{k}
-{\mid \Delta_{0}\mid^{2}\over G}-G\sum_{k}\rho_{k(0)}^{2};\nonumber\\
E_{jl}&=&2\sum_{k\ne j,l} \rho_{k(jl)}\epsilon_{k}
-{\mid \Delta_{jl}\mid^{2}\over G}\nonumber\\
& & -G\sum_{k\ne j,l}\rho_{k(jl)}^{2}+
\epsilon_{j}+\epsilon_{l};
\label{euri}
\end{eqnarray}
and $T_{k(\gamma)}$ are energy terms associated to single-particle states:
\begin{eqnarray}
T_{k(\gamma)}
&=&2\rho_{k(\gamma)}(\epsilon_{k}-\lambda_{\gamma})
-2G\rho_{k(\gamma)}^{2}\nonumber\\
& &+{\kappa_{k(\gamma)}\Delta_{\gamma}^{*}
+\kappa_{k(\gamma)}^{*}\Delta_{\gamma}\over 2}
\left({\rho_{k(\gamma)}^{2}\over \mid\kappa_{k(\gamma)}\mid^{2}}-1
\right).\nonumber
\end{eqnarray}
The following notations are used in Eqs. (\ref{system}):
\begin{eqnarray}
\Delta_{0}&=&G\sum_{k}\kappa_{k(0)};\nonumber\\
\Delta_{jl}&=&G\sum_{k\ne j,l}\kappa_{k(jl)};\nonumber\\
\kappa_{k(\gamma)}&=&u_{k(\gamma)}v_{k(\gamma)};\nonumber\\
\rho_{k(\gamma)}&=&\mid v_{k(\gamma)}\mid^{2};\label{notatii}\\
P_{\gamma}&=&\mid c_{\gamma}\mid^{2};\nonumber\\
S_{\gamma\gamma'}&=&c_{\gamma}c_{\gamma'}^{*};\nonumber
\end{eqnarray}
where $\rho_{k(\gamma)}$ are single-particle densities and 
$\kappa_{k(\gamma)}$ are pairing moment
components. $P_{\gamma}$ denote the probabilities to find the system in the
configurations $\gamma$. $S_{\gamma\gamma'}$ are moment components between two
configurations $\gamma$ and $\gamma'$ and have the property 
$\mid S_{\gamma\gamma'}\mid^{2}=P_{\gamma}P_{\gamma'}$. $\Delta_{\gamma}$ is
the gap parameter.  The values of $\rho_{k(\gamma)}$ and $P_{\gamma}$ 
 are reals.
The particle number conservation conditions $2\sum_{k}\rho_{k(0)}=2N$, 
$2\sum_{k\ne j,l}\rho_{k(jl)}=2N-2$ and $P_{0}+\sum_{j,l\ne j}P_{jl}=1$
are fulfilled by Eqs.  (\ref{system}).

\section{Application to fission processes}

A direct application of the system (\ref{system}) is related to the pair
breaking and the odd-even structure in fission fragment yields. In fission,
it is considered that the paired configuration 
is preserved until an interaction  breaks some
pairs in combination with the existence of a sufficient high excitation energy.
The odd-even structure in fission is explained usually within statistical
arguments, as for example in Refs. \cite{rejmund,avrig}.
Alternatively, the probability to
break a pair can be determined dynamically 
by taking into account only the interaction available in 
the avoided crossing regions within the present model.

\begin{figure}
\begin{center}
\resizebox{0.50\textwidth}{!}{
  \includegraphics{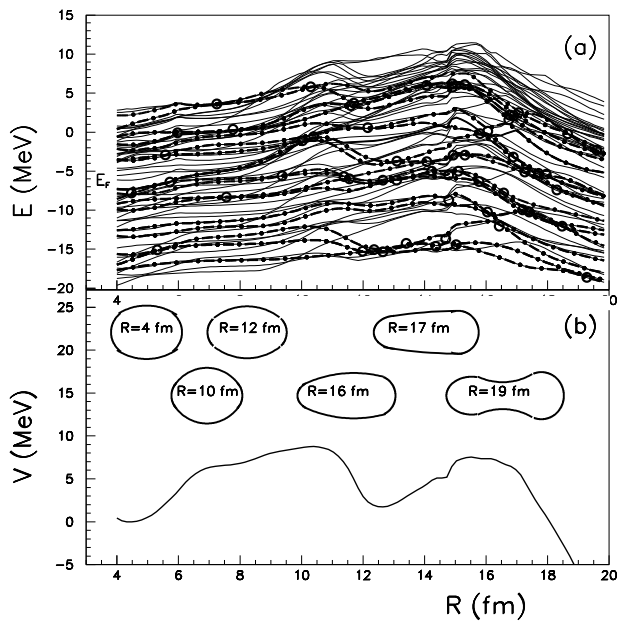}}
\caption
{(a) Proton level scheme as function of the inter-nuclear distance $R$
between the nascent fragments along the minimal action path for the
$^{234}$U fission. Only $N$=60 single-particle energies above and below
 the Fermi
level are plotted. The Fermi level $E_F$ is displayed with a thick full curve.
The selected  $\Omega$=1/2 levels ($\epsilon_{1},... \epsilon_{17}$)
are plotted with thick dot-dashed lines. Avoided level crossing regions
are marked with circles.
The maximum number of major shells used in calculations is 12. 
(b) The fission barrier as function of the internuclear distance $R$. 
Nuclear shapes corresponding to several special configurations are 
inserted in the plot. The values of the internuclear distances are also marked.
At $R\approx 4$ fm, the ground-state configuration of the parent is
found. 
}
\label{figura2}
\end{center}
\end{figure}

To solve the pair breaking equations (\ref{system}), the variations of
single-particle energies $\epsilon_{k}$
together with perturbations $h_{ij}$ must be
supplied. The simplest way to obtain these evolutions is to consider
a time-dependent mean field in which the nucleons move independently.
In most usual treatments of nuclear fission, the whole nuclear system
is characterized by some collective variables, which determine
approximately the behavior of many other intrinsic variables. 
The generalized coordinates vary in time leading to a split of
the nuclear system. 
The basic ingredient in such an analysis is a nuclear
shape parametrization.
In the following treatment, a nuclear 
shape parametrization
is given by two ellipsoids of different sizes smoothly joined by a third 
surface obtained \cite{mirea2}
by the rotation of a circle around the axis of 
symmetry. Five degrees of freedom characterize this parametrization:
the elongation given by the inter-nuclear distance $R$ between
the centers of the ellipsoids, the two deformations
of the nascent fragments, the mass asymmetry and the necking parameter.
Due to the axial symmetry, the good quantum numbers are the projections
of the intrinsic spin $\Omega$.

As specified in Ref. \cite{funny}, first of all, a calculation of 
the fission trajectory
in our five-dimensional configuration space, beginning with the ground-state
of the system up to the exit point of the barrier must be performed.
This can be done
by minimizing the action integral. For this purpose, two ingredients
are required: the deformation energy $V$ and the tensor of the effective mass.
The deformation energy was obtained in the frame
of the microscopic-macroscopic method \cite{nix}  by summing the liquid drop
energy with the shell and the pairing corrections.  
The macroscopic energy is obtained
in the framework of the Yukawa plus exponential model \cite{davies}
extended for binary systems with different charge densities
\cite{ejpa}. The Strutinsky microscopic corrections
 were computed on the basis of the Woods-Saxon 
superasymmetric two center 
shell model. The effective mass is computed within the cranking 
approximation \cite{rrp}. 
After minimization,
the dependences between the generalized coordinates $q_{i}$ ($i=1,...5$)
in the region comprised between the parent ground state configuration and the
exit point of the external fission barrier
supply the least action trajectory.
The ground-state corresponds to the lowest deformation
energy in the first well.  
The least action trajectory is obtained
within
a numerical method.
Details about the numerical procedure of minimization
and about the model can be found
in Refs. \cite{mirea,mirea2} and references therein. 
The resulting $^{234}$U fission barrier is plotted on
Fig. \ref{figura2} (b) as function of the distance between the
centers of the nascent fragments $R$. Some nuclear shapes
obtained along the minimal action trajectory are inserted in 
the plot. A realistic proton level scheme along the least action
trajectory is obtained within the superasymmetric Woods-Saxon 
two-center shell model \cite{mirea2}. This model gives the
single particle level diagrams by diagonalizing a Woods-Saxon potential,
corrected within spin-orbit and Coulomb terms, in the analytic eigenvalue basis
of the two center semi-symmetric harmonic model \cite{mnpa,gr}.
Other recipes to obtain the level scheme 
are related to the molecular orbital method \cite{diaz}.
The proton level scheme
is displayed in Fig. \ref{figura2} (a).

\begin{figure}
\begin{center}
\resizebox{0.50\textwidth}{!}{
  \includegraphics{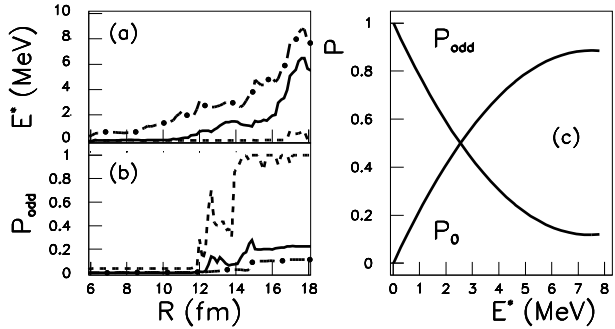}}
\caption{(a)  The average excitation energy $E^*$ as function
of the elongation $R$. The inter-nuclear velocities $\dot{R}$
are 10$^4$, 10$^{5}$ and
10$^6$ m/s for the dashed, full and dot-dashed lines, respectively.
(b) The  probability $P_{odd}$ of a seniority-two state 
with respect to the elongation $R$. 
The same line types and inter-nuclear velocities as in panel (a) are
used.
(c) The probabilities to obtain a seniority-zero state $P_0$ and a 
seniority-two state $P_{odd}$ as function
of the excitation energy $E^{*}$ of the fragments at the elongation $R$=20 fm.
}
\label{figura3}
\end{center}
\end{figure}

The Landau-Zener effect is produced only in the workspace spanned by
levels characterized by the same good quantum numbers. As mentionned,
due to the axial symmetry, the good quantum numbers are the 
projections of the intrinsic spin $\Omega$. Therefore,
among the $N$ states in the region near the Fermi surface, the
17 levels with spin projection $\Omega$=1/2 are selected. These
levels are plotted with thick dot-dashed lines in Fig. \ref{figura2} (a).
Because a pair creation or annihilation  is considered to be possible
only between adjacent levels, 16 seniority-two configurations are
constructed.
The initial values of quantities(\ref{notatii})
are obtained by solving the BCS equations for the ground states
of all $\gamma$ configurations. 
The average excitation energy of the 
seniority-zero state is computed
as in Ref. \cite{npa1,koonin}
with the relation
$E^{*}=(E_{0}-E_{0}^{0})+E_{n}^{*}$,
where $E_{0}^{0}$ is the value of the lowest energy state of any deformation
calculated within the BCS approach 
 and $E_{0}$
is obtained with Eqs. (\ref{euri}). Here, $E_{n}^{*}$ is the dissipation 
obtained within the same formalism
for the even neutron subsystem. 
The probability to obtain
a seniority-two state is simply
$P_{odd}=1-P_{0}$. 
The values of $E^{*}$ and $P_{odd}$ 
 are plotted in Fig. \ref{figura3} (a) and (b) as 
function of the elongation for some inter-nuclear 
velocities $\dot{R}$. In connection
with the shape of the barrier displayed in Fig. \ref{figura2}, it can be deduced
that the larger part of the odd-even yield is formed during the penetration
of the second barrier and the excitation energy increases merely in the same
region. 
Different constant values of the inter-nuclear 
velocity $\dot{R}$ ranging from 10$^{4}$ to
10$^{6}$ m/s were tested. These values can be translated in a time to penetrate
the barrier ranging in the interval $[1.4\times10^{-18},1.4\times10^{-20}]$ s.
In Fig. \ref{figura3} (c), the dependences of $P_0$ and $P_{odd}$
versus $E^{*}$ are displayed in the selected velocity
domain.  The results exhibit a clear decrease of $P_{odd}$ as function of $E^{*}$.
It is interesting to note that at zero excitation energy, the probability 
to find the system in a seniority-two state is practically one.
In cold fission, at very low excitation energies of the fragments,
the odd-even yields are always larger 
than the even-even ones \cite{sch,sch2,ham}.
The even-even fragmentation dominates at larger excitation energies
of the fragments, above 4-6 MeV.
It is a very strange phenomenon because in cold processes the system
doesn't possess enough energy to break a pair and because the penetrability
is hindered for odd-systems due to the specialization energies associated
to the two unpaired nucleons.
Up to now, the statistical explanation of this phenomenon involved 
some modifications of the
level densities for odd-even and even-even partitions \cite{avrig}
by according them within the deformations of the fragments
as function of the excitation
energies and the shell effects.
However,  
if one assumes that the odd-even effect 
in the fission fragments distribution
is strongly correlated to
the seniority-two state probability, 
this phenomenon can be alternatively 
explained by solving the coupled-channel system
of time-dependent pair breaking equations
as evidenced above. In this work, only 
the $\Omega$=1/2 subspace of the proton level diagram
is treated, but the same formalism can be applied
to other subspaces.

In conclusion, a new set of time-dependent coupled channel equations
derived from the variational principle is proposed
to determine dynamically the mixing 
between seniority-zero and seniority-two configurations. 
The essential idea is that the
configuration mixing is managed under the action of some inherent
low lying time dependent excitations produced in the avoided crossing
regions.
These equations were used to explain the odd-even effect in cold fission
processes. Only the radial coupling was used in the analysis, but it is
possible to extend the equations to 
take into account even the Coriolis coupling,
as in Ref. \cite{mirea3}. 
The main trends concerning the dependence of the odd-even
effect in fragments yields
versus the fragments excitation energy were reproduced. 
The formalism can be adjusted for other types of processes.
In this respect, the value of the interaction $h_{\gamma\gamma'}$
can also be the magnitude of other kind of interactions between
different configurations and the collective variables could be the
 amplitude of an external applied field as encountered
in the field of condensed matter.  The essential idea is the
inclusion in the energy functional of a time dependent low
perturbation by the mean of quasiparticle operators.

{\bf Acknowledgments}
This work was supported by the CNCSIS IDEI 512 contract
of the Romanian Ministry of Education and Research.



\end{document}